\documentclass[
    reprint,
    amsmath,
    amssymb,
    aps,
    superscriptaddress,
]{revtex4-2}

\usepackage{float}
\usepackage{bbm}
\usepackage{epsfig}
\usepackage{epstopdf}
\usepackage{pgfplots}
\usetikzlibrary{patterns}
\usepgfplotslibrary{groupplots}
\usepackage{amssymb}
\usepackage{amsmath}
\usepackage{scalefnt}
\usepackage{color}
\usepackage[title]{appendix}
\usepackage{hyperref}
\usepackage{qcircuit}
\usepackage{algorithm} 
\usepackage{algpseudocode}
\usepackage[capitalize]{cleveref}
\usepackage{braket,mleftright}
\usepackage{physics}
\usepackage{import}
\usepackage{dcolumn}
\usepackage{csquotes}
\usepackage{tikz}
\usetikzlibrary{calc}
\usepackage{tkz-graph}
\usepackage{graphicx} 
\usepackage{subfigure}
\usepackage{dsfont}
\usepackage{cleveref}

\newcommand{\be}{\begin{equation}}
\newcommand{\ee}{\end{equation}}
\newcommand{\bea}{\begin{eqnarray}}
\newcommand{\eea}{\end{eqnarray}}
\def\ket#1{\left| #1 \right\rangle}
\def\bra#1{\left\langle #1 \right|}

\newcommand{\NewName}{NoVa-ADAPT algorithm}

\begin{document}
\title{Non-Variational ADAPT algorithm for quantum simulations}

\author{Ho Lun Tang}
\affiliation{Department of Physics, Virginia Tech, Blacksburg, VA 24061, USA}
\affiliation{Virginia Tech Center for Quantum Information Science and Engineering, Blacksburg, VA 24061, USA}
\author{Yanzhu Chen}
\email{yanzhu.chen@fsu.edu}
\affiliation{Department of Physics, Virginia Tech, Blacksburg, VA 24061, USA}
\affiliation{Virginia Tech Center for Quantum Information Science and Engineering, Blacksburg, VA 24061, USA}
\affiliation{Department of Physics, Florida State University, Tallahassee, FL 32306, USA}
\author{Prakriti Biswas}
\affiliation{School of Electrical, Computer and Energy Engineering,
Arizona State University, Tempe, Arizona 85287, USA}
\author{Alicia B. Magann}
\affiliation{Quantum Algorithms and Applications Collaboratory, Sandia National Laboratories, Albuquerque, New Mexico 87185, USA.}
\author{Christian Arenz}
\affiliation{School of Electrical, Computer and Energy Engineering,
Arizona State University, Tempe, Arizona 85287, USA}
\author{Sophia E. Economou}
\affiliation{Department of Physics, Virginia Tech, Blacksburg, VA 24061, USA}
\affiliation{Virginia Tech Center for Quantum Information Science and Engineering, Blacksburg, VA 24061, USA}

\date{\today}

\begin{abstract} 
We explore a non-variational quantum state preparation approach combined with the ADAPT operator selection strategy in the application of preparing the ground state of a desired target Hamiltonian. In this algorithm, energy gradient measurements determine both the operators and the gate parameters in the quantum circuit construction.  
We compare this non-variational algorithm with ADAPT-VQE and with feedback-based quantum algorithms in terms of the rate of energy reduction, the circuit depth, and the measurement cost in molecular simulation. 
We find that despite using deeper circuits, this new algorithm reaches chemical accuracy at a similar measurement cost to ADAPT-VQE. Since it does not rely on a classical optimization subroutine, it may provide robustness against circuit parameter errors due to imperfect control or gate synthesis.
\end{abstract}

\maketitle
\section{Introduction}
 
Preparing the ground state of many-body Hamiltonians and finding the corresponding ground state energy is an important and challenging problem in physics and chemistry.
Classical simulation of such systems can require enormous computational resources due to the exponential scaling of the Hilbert space.
A potential alternative approach is to simulate the target problem using a quantum processor~\cite{Feynman1982}.
Here, the quantum phase estimation algorithm \cite{kitaev_quantum_1995, doi:10.1080/00268976.2011.552441} is a promising candidate for estimating ground state energies on quantum computers \cite{PhysRevLett.79.2586,PhysRevLett.83.5162}. However, to solve problems of a non-trivial size, its implementation requires deep quantum circuits and is expected to need fault-tolerant quantum computers~\cite{Preskill2018quantumcomputingin}.
In the Noisy-Intermediate-Scale-Quantum (NISQ) \cite{Preskill2018quantumcomputingin} era, quantum-classical hybrid algorithms provide a promising way to utilize more limited quantum computing resources, enabled by the power of classical computing~\cite{Peruzzo2014, Bauer2016, Yuan2019, Motta2020, Gomes2021}. 
One family of such algorithms is the variational quantum eigensolver (VQE) \cite{Peruzzo2014, McClean2016}, which targets the problem of preparing the ground state of a desired Hamiltonian. VQEs operate by preparing an ansatz state on a quantum device using a parameterized quantum circuit. Then, measurements of this state allow for estimating the expectation of the Hamiltonian under consideration. A classical computer is then used to variationally find the values of the quantum circuit parameters in order to minimize this Hamiltonian expectation value. Various VQE algorithms have been demonstrated in experiments~\cite{Peruzzo2014, PhysRevX.8.011021, Kandala2017, PhysRevA.95.020501, PhysRevX.8.031022}.

The performance of VQE algorithms relies on the quality of the variational ansatz.
An ideal ansatz can explore the Hilbert space containing the ground state with a minimal number of variational parameters and quantum gates.
The Adaptive Derivative-Assembled Problem-Tailored Variational Quantum Eigensolver (ADAPT-VQE) \cite{Grimsley2019} was introduced to construct ans{\"a}tze with this property. 
Instead of using a fixed ansatz, ADAPT-VQE grows the ansatz iteratively by adding the most effective operators.
In various numerical simulations, it has demonstrated promising improvement compared to fixed-circuit VQEs in terms of the number of variational parameters, the circuit depth, and the measurement cost \cite{Grimsley2019, Tang2021, anastasiou2022tetrisadaptvqe}.

Despite such efforts, the energy landscape for physical systems is typically complicated~\cite{Chen_2023_localminima}, containing sub-optimal solutions due to the non-convex nature of the optimization problem, which hinders the classical search for the global optimum. 
Although there exist many well-developed classical optimization algorithms, the optimization complexity inevitably increases with the system size and the number of variational parameters, and hence becomes a potential bottleneck of VQEs~\cite{PhysRevLett.127.120502}. This bottleneck can manifest as the need for a large number of function evaluations in the classical optimization procedure, which amounts to a large number of quantum circuits to be run. 
Furthermore, the accuracy of Hamiltonian expectation value estimates, limited by the inherent noise in quantum devices, has a significant impact on the classical optimization. 
For VQEs to offer practical utility, the classical and quantum resources associated with the variational procedure have to be kept at a reasonable level.

In this work, we introduce an adaptive quantum algorithm, referred to as the Non-Variational ADAPT (NoVa-ADAPT) algorithm, for preparing the ground state of a target Hamiltonian that does not require classical optimization, i.e., is non-variational.  
In the ansatz construction procedure of ADAPT-VQE, the energy derivatives with respect to a set of operators are measured, based on which the optimal operator is selected to be a generator of the ansatz.
In our \NewName, we select operators in the same way but do not optimize over the previous parameters that enter the state. Instead, we estimate the coefficient associated with the selected generator using the measured information from the operator selection process. This allows us to save the measurement cost associated with the classical optimization process, and can potentially reduce the impact of erroneous measurement results.
At the same time, selecting operators based on the energy gradient information allows us to save quantum and classical resources.

Previous work has considered the prospect of an optimization-free approach in both classical and quantum algorithms for quantum state preparation. 
Among them, the Anti-Hermitian Contracted Schr\"{o}dinger Equation (ACSE) method uses pre-determined operators to iterate the state in a classical algorithm~\cite{Mazziotti2006}.  
Another example is feedback-based quantum algorithms (FQAs), such as the Feedback-based ALgorithm for Quantum OptimizatioN (FALQON)~\cite{Magann_2022} and related works~\cite{Magann_2022,Magann2022_1,abdul2024adaptive,pexe2024using,rahman2024feedback,rahman2024feedback2,malla2024feedback,arai2024scalable,brady2024focqsfeedbackoptimallycontrolled,chandarana2024lyapunovcontrolledcounterdiabaticquantum,Larsen2023}, which utilize a fixed, repeating quantum circuit structure and operate without any classical optimization subroutine.   
Optimization-free quantum algorithms for ground state preparation have also been developed using the theory of Riemannian gradient flows~\cite{Schulte-Herbruggen2010, Wiersema2023, Magann_2023}. These latter strategies aim to minimize a cost function by moving in each adaptive step into the direction of the Riemannian gradient. Efficient implementation of this approach on quantum computers can be obtained by approximating the Riemannian gradient through $1$- and $2$-local Pauli operators~\cite{Wiersema2023}, or projecting the Riemannian gradient into a random direction~\cite{Magann_2023}. In fact, moving in each step into a random direction provably gives convergence to the ground state (almost surely) despite the existence of sub-optimal solutions~\cite{Malvetti2024} (i.e., in the form of saddle points). In the present work, we draw on these previous results and compare our algorithm's performance against this prior art.

In Sec.~\ref{sec:ADAPT-VQE} we briefly review ADAPT-VQE and its operator selection criterion, before describing the \NewName\ in Sec.~\ref{sec:NV-ADAPT}. We then review the feedback-based quantum algorithms that are most closely related in Sec.~\ref{sec:FB}.
In Sec.~\ref{sec:results}, we demonstrate NoVa-ADAPT with numerical simulations, and compare it to ADAPT-VQE and the previous feedback-based algorithms. 
We also investigate multiple approaches for estimating the coefficients in the non-variational approach and compare their performances in terms of the number of operators required in the state construction and the measurement cost.
Then in Sec.~\ref{sec:errors}, we explore the impact of the rotational error and the gradient measurement error on the performance, which are expected to be dominant in our algorithm.
Finally, we propose a hybrid algorithm of ADAPT-VQE and the non-variational ADAPT method in Sec.~\ref{sec:hybrid} for practical use in the near future.

\section{ADAPT-VQE} \label{sec:ADAPT-VQE}

ADAPT-VQE was introduced to dynamically construct a variational ansatz according to the problem Hamiltonian. 
It starts with an initial reference state and a predetermined operator pool $\{A_i\}$ consisting of a set of Hermitian operators $A_{i}$.
At the beginning of the $n$-th iteration, if the exponential of the operator $A_i$ is added to the ansatz, the energy expectation value will have the form
\be
    E_{A_i}(\theta) = \langle \psi^{(n-1)} \left| e^{-i\theta A_i} H e^{i\theta A_i} \right| \psi^{(n-1)} \rangle,
\ee
where $\ket{\psi^{(n-1)}}$ is the estimated ground state from the $(n-1)$-th optimization. 
ADAPT-VQE measures the first derivative of the energy with respect to the variational parameter associated with each candidate operator $A_i$ in the pool
\begin{align}
    \left.\frac{\partial E_{A_i}}{\partial \theta}\right|_{\theta=0} = i\langle \psi^{(n-1)} \left| \left[H,A_i\right]\right|\psi^{(n-1)}\rangle.
    \label{eq:1st_deriv}
\end{align}
The exponential of the operator $A^{(n)}$ with the largest derivative magnitude will be added to the previous ansatz $\ket{\psi^{(n-1)}}$ and form a new ansatz 
\begin{align}
    \ket{\psi^{(n)} (\theta_n, \dots, \theta_1)}=e^{i\theta_n A^{(n)}}\ket{\psi^{(n-1)} (\theta_{n-1}, \dots, \theta_1)}.
\end{align}
All the variational parameters $\{\theta_n, \dots, \theta_1\}$ in the updated ansatz are then varied to minimize the energy $E(\theta_n, \dots, \theta_1)$ by a classical computer with a chosen optimization algorithm.
Starting this optimization from the previous optimized state $\vert\psi^{(n-1)}\rangle$ corresponds to initializing the newly added parameter $\theta_n$ from $0$.  
This iteration is repeated until the norm of the measured gradient is smaller than a user-specified threshold.

The ADAPT-VQE algorithm has been studied extensively in numerical simulations~\cite{Grimsley2019, Tang2021, Shkolnikov2023avoidingsymmetry, Bertels2022, Warren2022, Romero2022, anastasiou2022tetrisadaptvqe, Grimsley2023, Long2024, VanDyke2024, Ramoa2024, Ramoa2024_1, Long2024, Farrell2024}. 
It has shown promising performance in finding the ground state with high accuracy and leads to much shallower quantum circuits than VQEs with a fixed ansatz. 
This is because the operator selection procedure adds the locally optimal operator to the ansatz at each iteration, which allows us to explore the Hilbert space efficiently. One can also incorporate symmetries of the Hamiltonian into the operators in the pool, further restricting the Hilbert space for the optimization procedure. 
Even for such favorable ans\"{a}tze, classically optimizing the variational parameters remains a challenge in the implementation of the algorithm, due to the large number of measurements required and the errors in the measurement results. 
In the next section, we investigate the strategy of abandoning the variational procedure and updating the state based on the information obtained in the operator selection procedure.

\section{NoVa ADAPT algorithm} \label{sec:NV-ADAPT}

Here, we introduce the \NewName.  
Similar to the original ADAPT-VQE algorithm, NoVa-ADAPT starts with a reference state and selects operators from a pre-determined operator pool based on the energy derivative.
We still denote the state obtained from the $(n-1)$-th iteration as $\ket{\psi^{(n-1)}}$, although no optimization is involved here. The energy derivatives will be of the same form as in Eq.~\ref{eq:1st_deriv}. At the $n$-th iteration, the operator $A^{(n)}$ with the largest derivative magnitude is used to update the state from $\ket{\psi^{(n-1)}}$ to 
\be
    \ket{\psi^{(n)}}=e^{i\eta_n A^{(n)}}\ket{\psi^{(n-1)}},
\ee
where $\eta_n$ is given by
\begin{align}
    \eta_n &= -\gamma \left.\frac{\partial E_{A^{(n)}}}{\partial \theta}\right|_{\theta=0} \\
    &= -\gamma \left.\frac{\partial }{\partial \theta}\Big\langle \psi^{(n-1)} \left| e^{-i\theta A^{(n)}} H e^{i\theta A^{(n)}} \right|\psi^{(n-1)}\Big\rangle \right|_{\theta=0}. \label{eq:eta}
\end{align} 
The factor $\gamma$ is a real number controlling the magnitude of the update. Although it may look similar to the ADAPT-VQE ansatz, here $\ket{\psi^{(n)}}$ is \textit{not} a variational ansatz since $\eta_n$ has a fixed value. We now denote the energy expectation value obtained after the $n$-th iteration as $E^{(n)} \equiv \langle\psi^{(n)}\vert H \vert\psi^{(n)}\rangle$ for convenience.

As the gradient is evaluated at zero parameter value, a sufficiently small $\gamma$ can ensure that the energy is lowered, i.e., $E^{(n)}-E^{(n-1)}\leq0$.  
More specifically, based on a descent lemma for first-order optimization algorithms~\cite{Magann_2023, Beck_opt}, it was shown that the energy reduction can be lower bounded by 
\begin{align}
    E^{(n-1)}-E^{(n)} \geq \frac{1}{8\lVert H\rVert_2 \lVert A^{(n)} \rVert_2^2}\left( \left.\frac{\partial E_{A^{(n)}}}{\partial\theta}\right|_{\theta=0}\right)^2,
\end{align}
by choosing 
\begin{align}
    \gamma = \frac{1}{4\lVert H\rVert_2 \lVert A^{(n)}\rVert_2^2},
    \label{eq:gamma_bound}
\end{align}
where $\lVert M\rVert_2$ is the spectral norm of operator $M$ and $A^{(n)}$ is the operator added to the state at this iteration~\cite{Magann_2023}.
Although this choice ensures a non-zero energy reduction given a non-zero energy derivative, it limits the effect of each operator added to the state, thus increasing the number of operators required to reach the ground state.

To estimate the optimal value of $\gamma$ to minimize the energy, we expand the energy reduction to the second order, 
\begin{align}
\label{eq:taylor}
    E^{(n)}-E^{(n-1)} \approx \eta_n \left.\frac{\partial E^{(n)}}{\partial \eta_n} \right|_{\eta_n=0} + \frac{\eta_n^2}{2}\left.\frac{\partial^2 E^{(n)}}{\partial \eta_n^2} \right|_{\eta_n=0},
\end{align}
and find the stationary point
\begin{align}
    \gamma^* = -\left(\left.\frac{\partial^2 E^{(n)}}{\partial \eta_n^2}\right|_{\eta_n=0}\right)^{-1}.
    \label{eq:2nd_deriv}
\end{align}
In practice, this requires additional measurements for the second derivative, but at each iteration we only need to evaluate it for the selected operator $A^{(n)}$ so the cost does not grow with the size of the operator pool.

The non-variational approach saves the sampling cost associated with the optimization process. On one hand, this can potentially reduce the number of measurement samples required by the algorithm as the measurement cost in ADAPT-VQE is dominated by the optimization process. On the other hand, to generate each measurement sample, each quantum circuit evaluation is subject to noise, and eliminating the optimization based on noisy measurements may reduce the impact of errors. 
We remark that the noise may have different effects on the operator selection process and the optimization process of ADAPT-VQE, which requires further studies in the future.

\section{Gradient-based feedback} \label{sec:FB}
The strategy of preparing the ground state based on the gradient estimate in an optimization-free way serves as the basis of feedback-based quantum algorithms (FQAs)~\cite{Magann_2022, Larsen2023} and the randomized adaptive quantum state preparation algorithm~\cite{Magann_2023}. While the \NewName\ updates the parameter in the same way, it differs from the previous algorithms in the operator selection. FQAs were inspired by quantum control protocols and use the terms in the Hamiltonian as generators for each iteration in the circuit, whereas the randomized adaptive algorithm gains robustness from random operators as generators. 
A classical algorithm for calculating reduced density matrices in chemistry adopted a similar way of iterating the state, with a linear combination of fermionic operators as the generator~\cite{Mazziotti2006}. A quantum-classical hybrid version of this algorithm turns the transformation on the state into a quantum circuit with Trotterization~\cite{Smart2021}. We briefly review these closely related approaches and compare the simulated performances in order to study the importance of operators.

\subsection{Feedback-based quantum algorithms}

The FQA procedure for preparing molecular ground states starts with separating the Hamiltonian into two parts,
\begin{equation}
    H=H_1+H_2,
\end{equation}
where the first part collects the single-body operators,
\begin{equation}
    H_1 = \sum_{q}h_{pq}a^\dagger_{p}a_{q},
\end{equation}
and the second part includes all the two-body operators in the Hamiltonian
\begin{equation}
    H_2=\frac{1}{2}\sum_{pqrs}a^\dagger_p a^\dagger_q a_r a_s.
\end{equation}
The FQA circuit of $n$ layers can be written as 
\begin{equation}
    U_{\rm FQA}^{(n)}(\vec{\beta})=\tilde{U}_1(\beta_n\Delta t) \tilde{U}(\Delta t)\cdots \tilde{U}_1(\beta_1\Delta t) \tilde{U}(\Delta t),
\end{equation}
where $\tilde{U}(\Delta t)$ and $\tilde{U}_1(\beta_k\Delta t)$ are the Trotterized versions of $e^{-i H \Delta t}$ and $e^{-i\beta_k H_1 \Delta t}$, respectively. The parameter $\beta_n$ is evaluated with the state obtained in the last iteration,
\begin{equation}
    \beta_n = - i\langle \psi^{(n-1)} \left| [H_1,H_2]\right| \psi^{(n-1)}\rangle,
\end{equation}
where $\ket{\psi^{(n-1)}}=U_{\rm FQA}^{(n-1)}(\vec{\beta})\ket{\psi^{(0)}}$. Conventionally, $\ket{\psi^{(0)}}$ is taken to be the ground state of $H_1$ in FQAs. In the material below, we deviate from this convention. We also remark that the time step, $\Delta t$, has to be chosen separately, which plays the same role as the $\gamma$ parameter in Eq.~\eqref{eq:eta}. One option is to use the lower bounding value introduced in Eq.~\eqref{eq:gamma_bound}, substituting $H_1$ for $A^{(n)}$.

\subsection{\label{sec:Randomized adaptive quantum algorithm}Randomized adaptive quantum algorithm}
 
The randomized adaptive quantum algorithm~\cite{Magann_2023} dynamically constructs the quantum circuit through randomly selecting an operator in each adaptive step, thereby achieving convergence to the ground state almost surely~\cite{Malvetti2024}. 
We consider here randomization by constructing the operators $\{A^{(n)}\}$ in two different ways: 
\begin{itemize}
    \item Sampling a random operator from a pool of operators and measuring the gradient of the cost function. The sampled operator is added to the state as along as it gives a non-zero gradient.
    \item Conjugating a fixed traceless operator $A$ by a random unitary transformation $V_n$ as
    \begin{equation}
        A^{(n)}=V_n^\dagger A V_n.
    \end{equation}
    The random unitary transformation $V_n$ can be sampled from a unitary 2-design that can be efficiently approximated by the construction in Ref.~\cite{Nakata_2017}.
\end{itemize}

\subsection{\label{sec:ACSE}Anti-Hermitian Contracted Schr\"{o}dinger Equation}
The ACSE method~\cite{Mazziotti2006} approaches the ground state of a molecular Hamiltonian $H$ iteratively, acting on the previous state $\ket{\psi}$ with $e^{\epsilon S}$, where $\epsilon$ is an infinitesimal step size and the anti-Hermitian operator $S$ contains up to two-body fermionic operators
\be
    S = \sum_{p,r} {^1S}_{r}^{p} a_p^\dagger a_r + \sum_{p,q,r,s} {^2S}_{r,s}^{p,q} a_p^\dagger a_q^\dagger a_ra_s.
\ee 
The matrix elements
\begin{align}
    & {^1S}_{r}^{p} = \bra{\psi} [a_p^\dagger a_r, H] \ket{\psi}, \nonumber \\
    & {^2S}_{r,s}^{p,q} = \bra{\psi} [a_p^\dagger a_q^\dagger a_ra_s, H] \ket{\psi}, 
    \label{eq:acse_commutator}
\end{align}
are selected to reduce the energy expectation value in an Euler's method. 
By Trotterizing $e^{\epsilon S}$, one can compile it as a quantum circuit and turn this classical algorithm into a quantum-classical hybrid algorithm, the quantum contracted-eigenvalue-equation solver~\cite{Smart2021, Boyn2021}. This may produce a similar circuit structure to the \NewName\ with a pool containing 1-body and 2-body fermionic operators, as the matrix elements of $S$ are the same as the energy gradients (up to a factor of $i$) and $\epsilon$ plays the same role as $\gamma$ in the \NewName. However, the states constructed by the two algorithms are very different. In this algorithm, a sequence of exponentials $\prod_k e^{i\alpha_kO_{k}}$ of all 1-body and 2-body fermionic operators $\{O_k\}$ is appended to the state at each iteration, with $\{\alpha_k\}$ calculated for this iteration. The effect depends on the choice of ordering in the Trotterization. In contrast, the NoVa-ADAPT approach only selects the one term with the largest magnitude of the coefficient. 
The difference between the circuit structures is analogous to the difference between the unitary coupled-cluster single and double (UCCSD) ansatz~\cite{Peruzzo2014} and the ADAPT-VQE ansatz~\cite{Grimsley2019}. Moreover, even if the selected operators coincide with the ordered sequence $\{O_k\}$, the update parameters are calculated from many iterations of the state, and will be different from $\{\alpha_k\}$. Because each iteration of the quantum contracted-eigenvalue-equation solver involves a sequence of terms, we expect the \NewName\ to lead to more compact quantum circuits.

\begin{figure}
    \centering
    \begin{subfigure}
        \centering
        \includegraphics[width=0.45\textwidth]{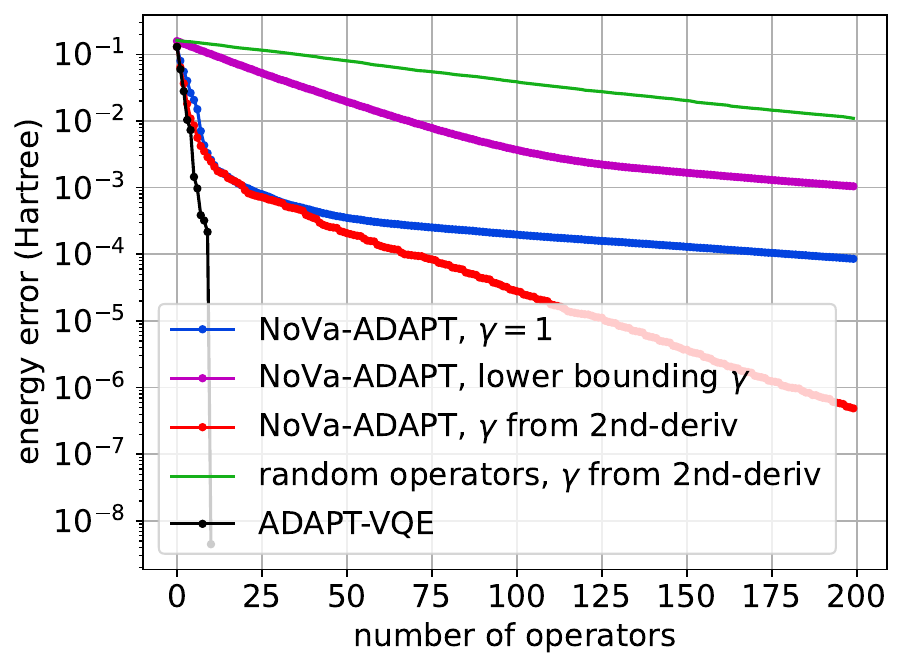}
        \label{fig:nop}
    \end{subfigure}
    \hfill
    \begin{subfigure}
        \centering
        \includegraphics[width=0.45\textwidth]{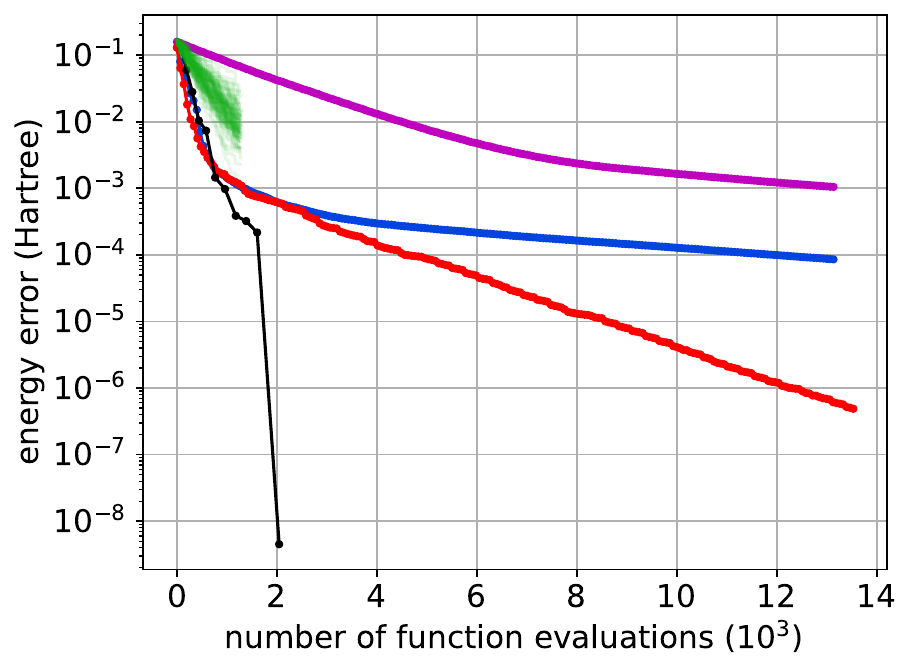}
        \label{fig:nfev}
    \end{subfigure}
    \caption{Results for $H_4$ molecule of bond length $1.5$\AA\ with the spin-adapted fermionic operator pool~\cite{Grimsley2019}. Energy error is plotted as a function of the number of operators added to the state (upper panel), and as a function of the number of function evaluations needed (lower panel). Blue curves show the results of the \NewName\ with constant $\gamma=1$. Green curves show the results of constructing the state with operators randomly picked from the pool with $\gamma$ calculated from the second derivative using Eq.~\ref{eq:2nd_deriv}, sampled from 100 runs. Magenta curves show the results of the \NewName\ with lower bounding $\gamma$ from Eq.~\ref{eq:gamma_bound}. Red curves show the results of the \NewName\ with $\gamma$ calculated from the second derivative using Eq.~\ref{eq:2nd_deriv}. Black curves show the results of ADAPT-VQE, where classical optimization is implemented. }
    \label{fig:hf}
\end{figure}

\begin{figure}
    \centering
    \includegraphics[width=0.45\textwidth]{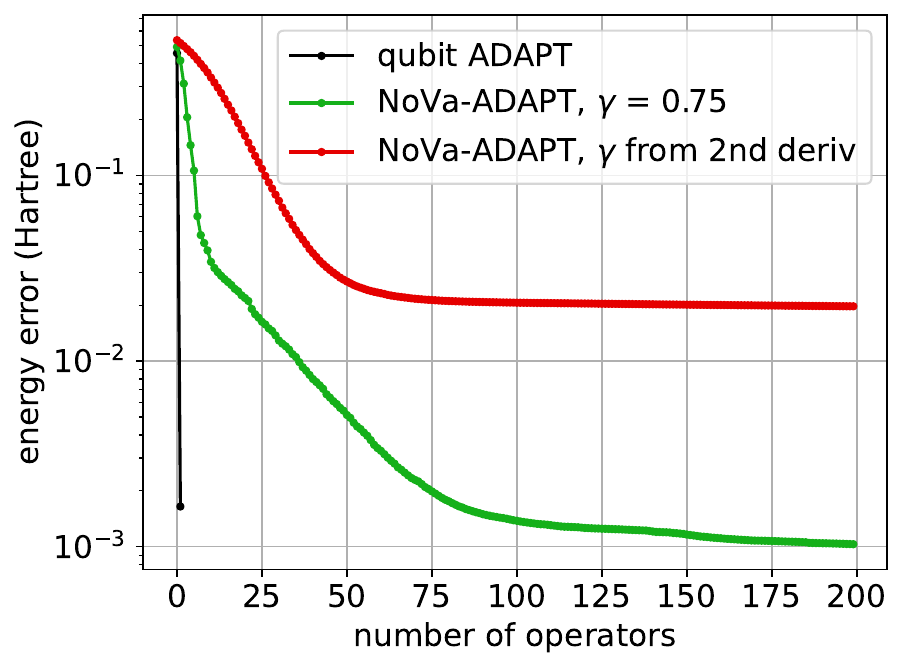}
    \caption{Results for $H_4$ molecule of bond length $3$\AA\ with the qubit operator pool~\cite{Tang2021}. Energy error as a function of the number of operators added to the state. The black curve shows the result of qubit-ADAPT-VQE, the green curve shows the result of the \NewName\ with constant $\gamma =0.75$, and the red curve shows the result of the \NewName\ with $\gamma$ calculated with the second derivative of energy using Eq.~\ref{eq:2nd_deriv}.}
    \label{fig:qubit_pool}
\end{figure}

\section{Results} \label{sec:results}
We simulate the performance of the proposed algorithm for finding the ground state of molecular Hamiltonians, with the Hartree-Fock state as the reference state. The Hartree-Fock state is a product state calculated with a mean-field theory approach to approximate the ground state. In our simulations, we take the $\text{H}_4$ molecule in the STO-3G basis with a bond distance of $1.5$ \AA\ as our main example. The Hamiltonian consists of fermionic creation and annihilation operators,
\begin{equation}
    H = \sum_{ij} h_{ij} a^\dagger_i a_j + \frac{1}{2}\sum_{ijkl} h_{ijkl} a^\dagger_i a^\dagger_j a_k a_l,
\end{equation}
where $\{i,j,k,l\}$ index the $8$ spin-orbitals we consider.
For ADAPT algorithms, we start from the restricted Hartree-Fock state, and use an operator pool containing the spin-adapted fermionic operators, which preserve the number of electrons, spin polarization $S_z$, and total spin $S^2$~\cite{Grimsley2019}. They consist of single excitation operators:
\begin{equation}
    \tau_1 \propto \ket{\uparrow}_a\bra{\uparrow}_b+\ket{\downarrow}_a\bra{\downarrow}_b-{\rm h.c.},
\end{equation}
and double excitation operators:
\begin{equation}
    \begin{split}
        \tau_{2,T} \propto & \ket{T,1}_{pq}\bra{T,1}_{rs}+\ket{T,-1}_{pq}\bra{T,-1}_{rs}  \\
    &+ \ket{T,0}_{pq}\bra{T,0}_{rs} - {\rm h.c.} \\
      \tau_{2,S} \propto &\ket{S,0}_{pq}\bra{S,0}_{rs}-{\rm h.c.},
      \label{spin_adapt}
    \end{split}
\end{equation}
where $\{a,b,p,q,r,s\}$ are spatial orbitals and $T$ and $S$ refer to triplet and singlet states formed by $p,q$ or $r,s$. 

In our simulations, we map these operators to Pauli operators using the Jordan-Wigner transformation, where 8 qubits are required. The expression of the Hamiltonian in terms of Pauli operators is given in Appendix.~\ref{app:H}.  
As both the Hamiltonian and the state are represented using fermionic operators, the algorithm and our simulation results are independent of the qubit encoding. 

We use the energy error, which is the difference between the energy produced by the algorithms and the energy from exact diagonalization, to measure the quality of the ground state. We use the number of function evaluations in each algorithm as a proxy for the measurement budget, where estimating either the energy or the gradient counts as one function evaluation.
The classical optimization in ADAPT-VQE simulations are carried out with Broyden–Fletcher–Goldfarb–Shanno algorithm (BFGS) algorithm and the Hessian approximated by BFGS is recycled for the next ADAPT iteration, this hessian recycling scheme introduced in \cite{Ramoa2024} can reduce the measurement cost in ADAPT-VQE. 
Fig.~\ref{fig:hf} compares the \NewName\ with ADAPT-VQE and the randomized approach. The state construction with operators randomly selected from the pool (green) needs the largest number of operators to reach a given energy accuracy. This is expected since the operators are not selected based on the energy gradient information. In terms of the measurement cost, it performs better than the \NewName\ with the lower-bounding $\gamma$ (magenta), which provides the lowest update rate for the non-variational state construction. 
The \NewName\ exhibits similar behaviors at early stages with constant $\gamma=1$ (blue) and with $\gamma = \gamma^*$ as in Eq.~\ref{eq:2nd_deriv}(red). Eventually, estimating the value of $\gamma$ based on the second derivative leads to a better energy accuracy, as this provides more information about the optimization landscape. 
The ADAPT-VQE simulation finds the ground state with 11 operators which is much lower than the \NewName. 
However, if we compare the number of function evaluations required in the algorithm (Fig.~\ref{fig:hf}(b)), the ADAPT-VQE performance is comparable to the \NewName\ with constant $\gamma=1$ and with $\gamma = \gamma^*$ until both algorithms reach chemical accuracy ($\approx 0.0016$ Hartree). 

On the other hand, ADAPT-VQE can terminate prematurely after converging to a local minimum in the energy landscape, in which case the \NewName\ may provide an alternative approach to avoid this situation. We pick one such case in Fig.~\ref{fig:qubit_pool}, the $H_4$ molecule with bond distance $3$\AA\, which is a strongly correlated Hamiltonian as the electrons are highly non-local.
Using the qubit operator pool~\cite{Tang2021}, which consists of the simplified individual Pauli operators in the Jordan-Wigner mapping of one-body and two-body fermionic operators, ADAPT-VQE finds a local minimum after adding the second operator. 
For this Hamiltonian, the \NewName\ with $\gamma$ based on the second derivative also reaches a region where the energy reduction slows down exponentially.
With a constant $\gamma=0.75$, however, the \NewName\ can lower the energy down to chemical accuracy, beyond the point where ADAPT-VQE was trapped. 
This is because the constant $\gamma$ strategy only learns about the first derivative of the energy landscape and thus avoids the local minimum, while the update based on the second derivative leads to the sub-optimal trap.

\subsection{Random initial states}

The performance of classical optimization subroutine can strongly depend on the initial parameter values.
For some Hamiltonians, the classically calculated reference state has a very small overlap with the ground state, which may hinder VQE's performance, requiring many more iterations to achieve convergence.
In this case, state preparation with randomly selected operators may be favorable as the initial state cannot introduce bias in this case.
Here, we compare ADAPT-VQE, the \NewName, and the randomized adaptive algorithm on the same $\text{H}_4$ molecule with a bond distance of $1.5$ \AA\ and a spin-adapted fermionic operator pool, but from randomly sampled initial states.

\begin{figure}
    \centering
    \begin{subfigure}
        \centering
        \includegraphics[width=0.45\textwidth]{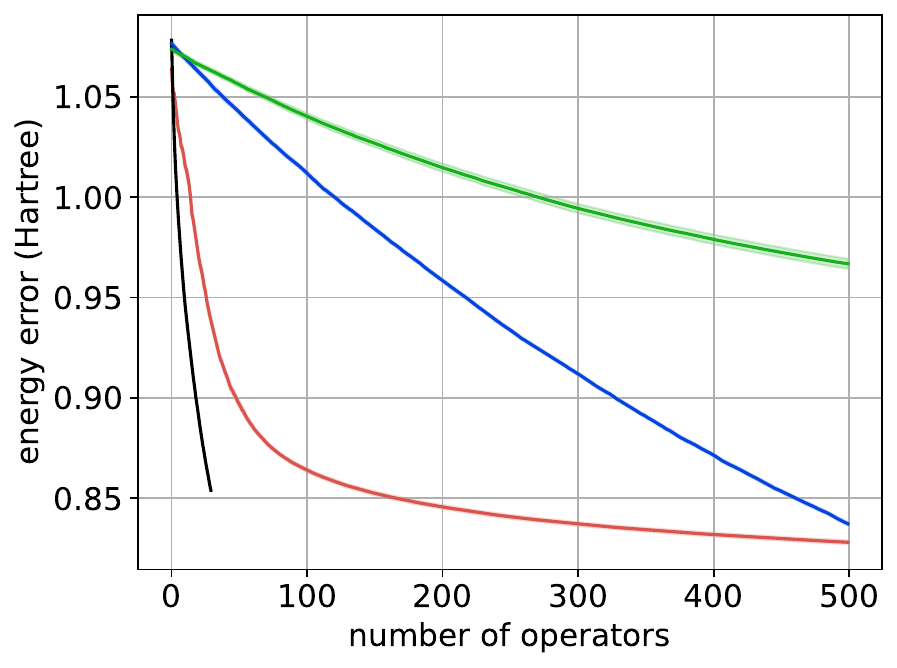}
    \end{subfigure}
    \hfill
    \begin{subfigure}
        \centering
        \includegraphics[width=0.45\textwidth]{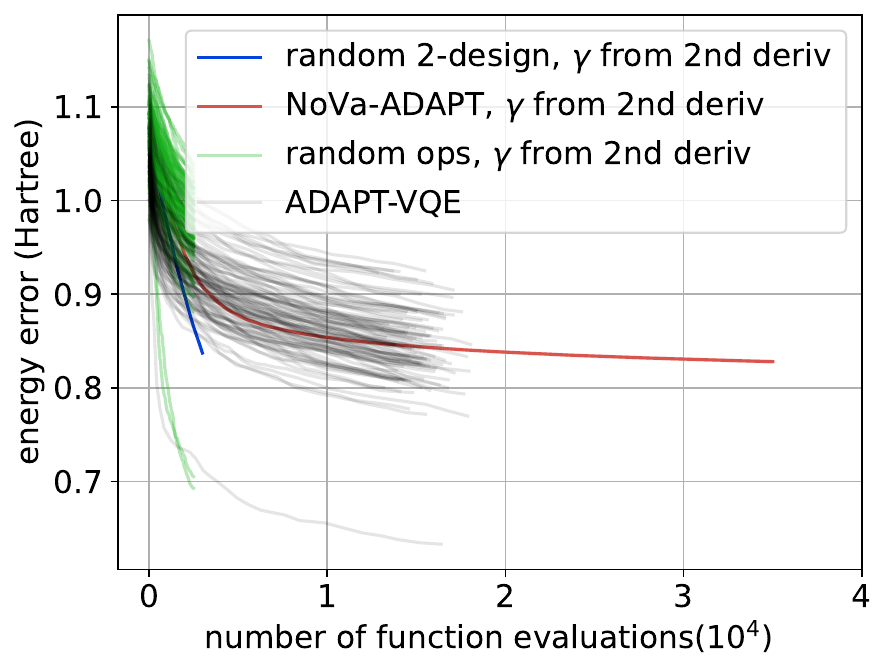}
    \end{subfigure}
    \caption{Results for $H_4$ molecule of bond length $1.5$\AA\ with 100 random initial states, obtained by ADAPT-VQE (black), the \NewName\ (red), randomly sampling operators from the pool (green), and the 2-design randomized adaptive method (blue). In the latter three non-variational approaches, $\gamma$ is computed with the second derivative. (Upper panel) Energy error is plotted as a function of the number of operators added to the state, averaged over random initial states. (Lower panel) Energy is plotted as a function of the number of function evaluations. Black curves and green curves show the results of 100 random initial states. Green curves include the measurements associated with the operators rejected by the algorithm (i.e., with zero energy derivative). The red curve shows the average over 100 random initial states. The blue curve shows the average of 100 samples for different initial states and different random 2-design circuits.}
    \label{fig:rand_init}
\end{figure}

The random initial states are obtained by sampling from a unitary $V$ from 2-design and applying $V$ to the Hartree-Fock state.
For all the methods except the original ADAPT-VQE, the parameters are estimated using the second derivative with Eq.~\eqref{eq:2nd_deriv}.
From Fig.~\ref{fig:rand_init}, we see that none of the methods can reach the ground state within a reasonable number of function evaluations.
For ADAPT-VQE (black), even if the simulation stops at the 30th iteration, the number of function evaluations has already exceeded that for the \NewName\ (red) which contains 500 operators. 
Adding randomized unitaries to the state (blue and green) lowers the energy by a smaller amount for a given number of operators, but the number of measurements associated with a given energy reduction is much lower than that for ADAPT-VQE and the \NewName. 
In this case, the random initial state breaks the symmetries in the problem whereas the operators in the ADAPT pool preserve the symmetries, which are therefore inefficient in approaching the symmetric ground state.

\subsection{Comparison to FQA and the quantum ACSE method}

\begin{figure}
        \centering
        \begin{subfigure}
            \centering
            \includegraphics[width=0.45\textwidth]{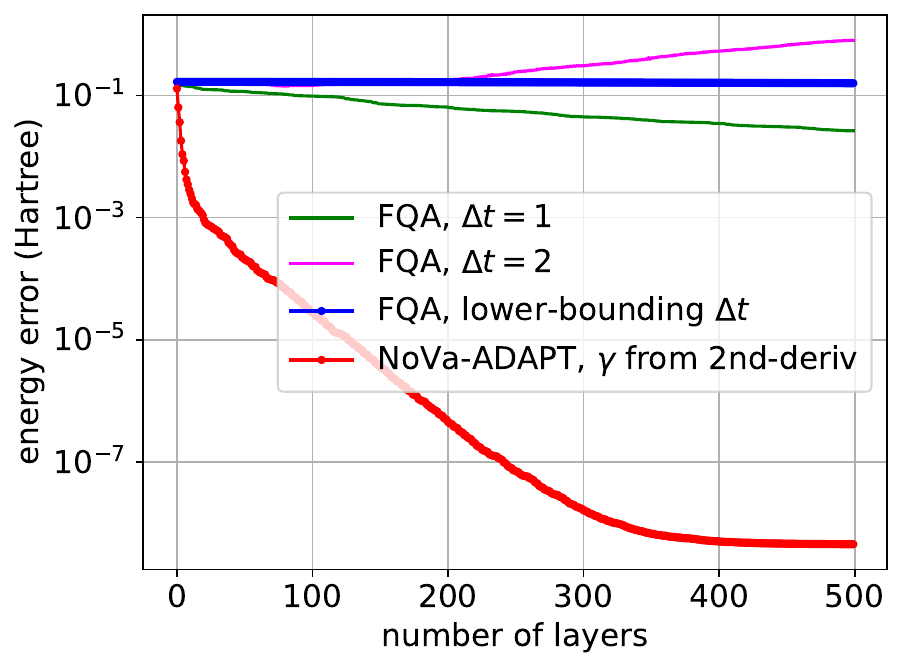}
        \end{subfigure}
        \hfill
        \centering
        \begin{subfigure}
            \centering
            \includegraphics[width=0.45\textwidth]{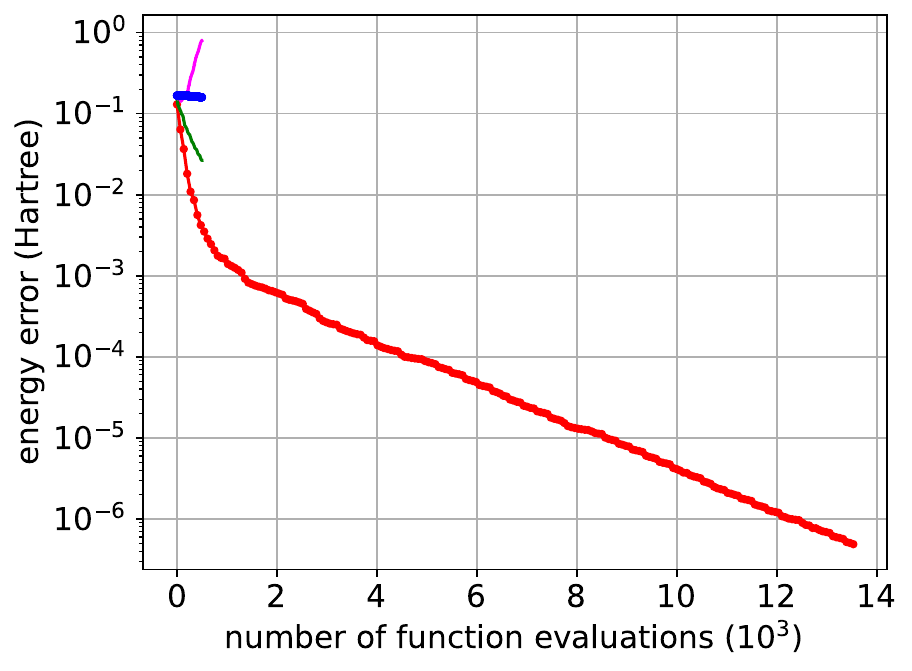}
        \end{subfigure}
    \caption{Results for $H_4$ molecule of bond length $1.5$\AA\ using the \NewName\ and FQA. Energy error as a function of the number of iterations/layers in the state(upper panel) and as a function of the number of function evaluations needed (lower panel). The red curve shows the result of the \NewName\ with $\gamma$ computed with second derivative. The blue, green, and pink curves show the results of FQA with the lowering bounding $\Delta t$, $\Delta t=1$, and $\Delta t=2$, respectively. For FQA, each layer in the state consists of two unitaries, i.e., $\tilde{U}_1(\beta_k\Delta t)$ and $\tilde{U}(\Delta t)$.}
    \label{fig:fqa}
\end{figure}

\begin{figure}
    \centering
    \begin{subfigure}
        \centering
        \includegraphics[width=0.45\textwidth]{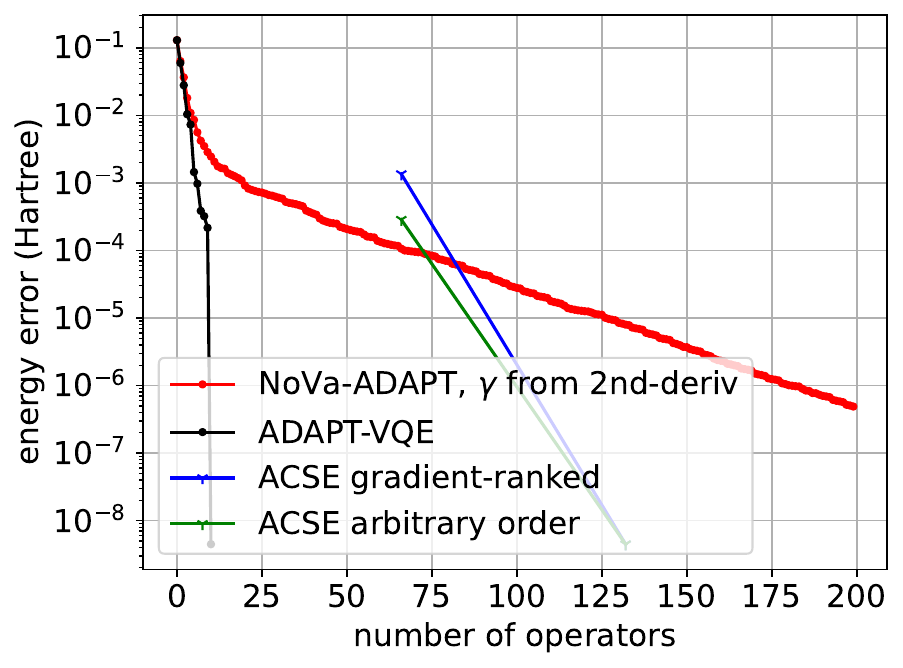}
    \end{subfigure}
    \hfill
    \centering
    \begin{subfigure}
        \centering
        \includegraphics[width=0.45\textwidth]{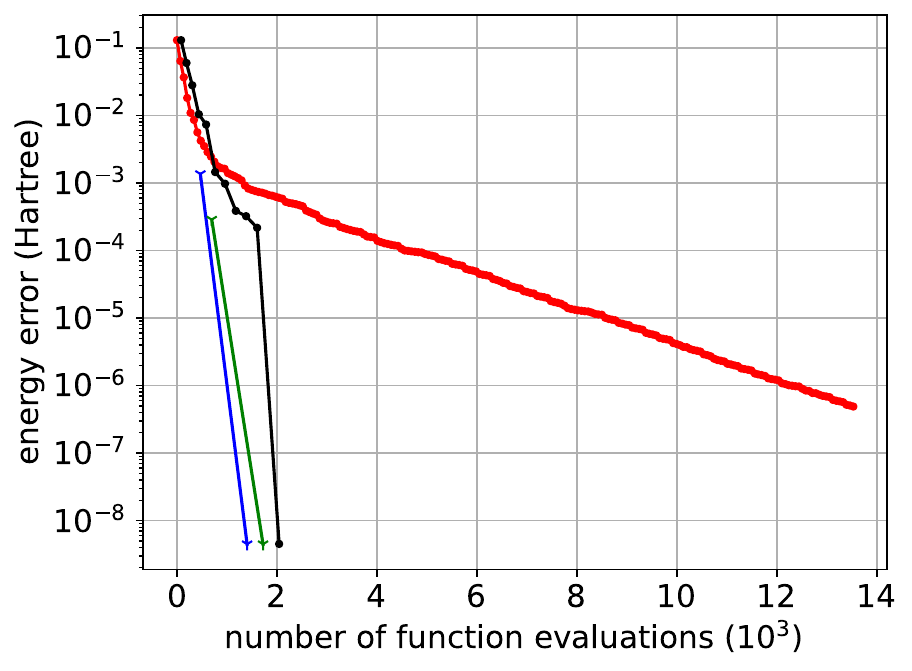}
    \end{subfigure}
    \caption{Results for $H_4$ molecule of bond length $1.5$\AA\ using the \NewName\ and the quantum ACSE method. Energy error is plotted as a function of the number of operators added to the state (upper panel) and as a function of the number of function evaluations needed (lower panel). The red curve shows the result of the \NewName\ with $\gamma$ computed with second derivative. The green and blue curves show the results of the quantum ACSE method with an arbitrary Trotter ordering and a Trotter ordering according to the energy gradient ranking, respectively.}
    \label{fig:acse}
\end{figure}

In Fig.~\ref{fig:fqa}, we see the \NewName\ with $\gamma$ computed from the 2nd-derivative performs substantially better than an FQA with various values of $\Delta t$ in terms of the number of iterations. It is worth noting that in the FQA, each layer consists of two unitaries, $\tilde{U}_1(\beta_k\Delta t)$ and $\tilde{U}(\Delta t)$. Each unitary is a product of a series of terms, since both $H_1$ and $H$ contain multiple single-body or two-body operators. Therefore, even with the same number of iterations, the circuit depth of FQA exceeds that of the \NewName. 
For the FQA, the energy reduction rate increases with $\Delta t$ for $\Delta t=1$ and below, with $\Delta t=2$ outside the range of energy reduction. 

In Fig.~\ref{fig:acse}, we compare the \NewName\ and the quantum ACSE method, where all the operators in the pool of the former appear in one iteration of the latter. In the quantum ACSE method, the value of $\epsilon$ is optimized at each iteration and the optimized value is fixed for subsequent iterations~\cite{Smart2021}. We use BFGS for this optimization, although in principle any optimizer can be used. Different choices of ordering can be made for the operators in each Trotter step. We show the results of an ordering following an arbitrary labelling of the operators (green), and of an ordering according to the magnitudes of the energy gradients, from the largest to the smallest (blue). The two ordering lead to similar performances in terms of the energy accuracy and the measurement budget. Since the \NewName\ adds only the most relevant operator at each iteration, it reaches chemical accuracy with a smaller number of operators than the quantum ACSE method, leading to more compact quantum circuits. 
The quantum ACSE method is more economical in the number of measurements required to reach the chemical accuracy, as all the operators in the pool are added in each iteration and therefore fewer iterations are needed.

From the comparison to FQA and the quantum ACSE method, we show that selecting the most energy-relevant operators improves the circuit depth of the feedback-based state preparation starting from the Hartree-Fock state.

\section{Noise Robustness} \label{sec:errors}

In practice, quantum devices are subject to errors due to imperfect control as well as interactions with the environment. 
We model the impact of errors on the performance of the algorithms studied in this work in two ways: errors in the variational or update parameters, which we refer to as rotational errors, and errors in the gradient measurements, as these directly affect the quantum circuit used for state preparation.
In the near term, without the protection of error correction, the rotational error stems mainly from the imprecise control of gate parameters. Even in the fault-tolerant regime, the implementation of the algorithm is not completely free of the rotational error. At the logical level, the actual gate parameter can still deviate from the desired value due to fault-tolerant gate synthesis with a finite overhead~\cite{Kitaev1997, Dawson2005, Ross2016}. Since the update parameter does not have to be a precise value for the energy to decrease, we anticipate the \NewName\ to be more robust to the rotational error than ADAPT-VQE. On the other hand, both the variational and the non-variational algorithms in this work select operators based on gradient estimates, which suffer from statistical uncertainty as well as decoherence. The non-variational approach also relies on these estimates to provide update parameters. Consequently, gradient measurement errors may have a larger impact on its performance. To investigate the robustness to these two error types, we perform simulations with modeled effective errors. The largest number of operators selected by a given algorithm is set to be $200$, beyond which the simulation is terminated by hand.

\begin{figure}
    \centering
    \begin{subfigure}
        \centering
        \includegraphics[width=0.45\textwidth]{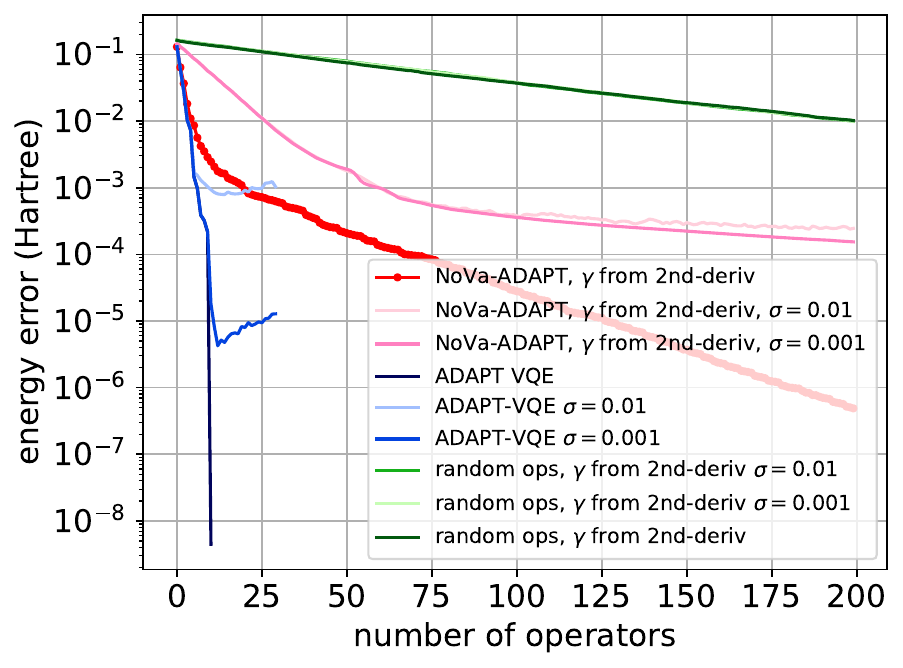}
    \end{subfigure}
    \begin{subfigure}
        \centering
        \includegraphics[width=0.45\textwidth]{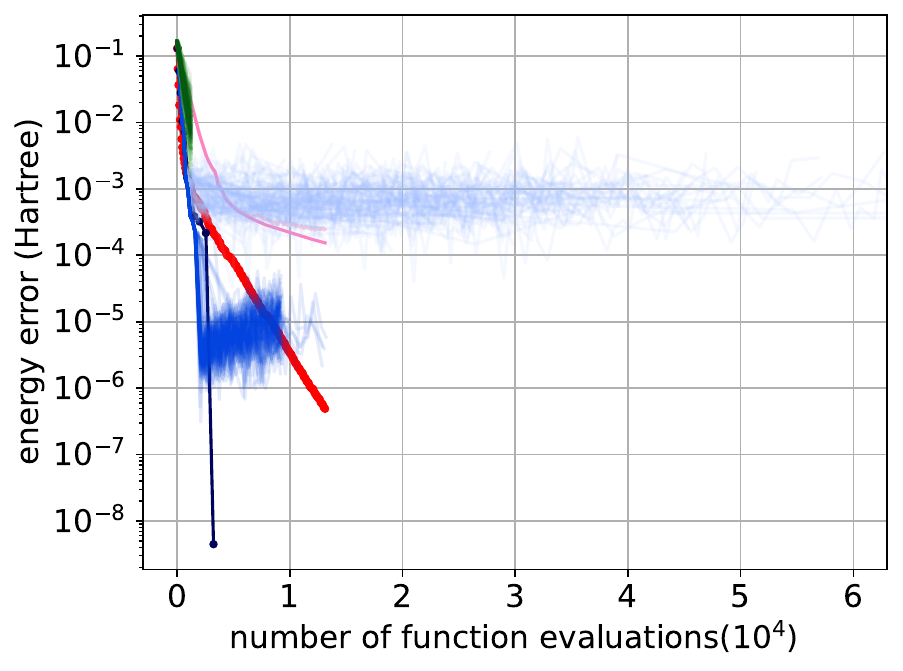}
    \end{subfigure}
    \caption{Results for $H_4$ molecule of bond length $1.5$\AA\ with rotational errors. Energy error is plotted as a function of the number of operators added to the state (upper panel), and as a function of the number of function evaluations (lower panel). Blue curves show the results of ADAPT-VQE. Red curves show the results of the \NewName\ with $\gamma$ computed from the second derivative. Green curves show the results of randomly picked operators from the pool with $\gamma$ computed from the second derivative. All results shown here are based on 100 circuit realizations, either as individual realizations or as the average.}
    \label{fig:rot_error}
\end{figure}

The rotational error is simulated by adding a random value to the optimized parameters at the end of each ADAPT iteration,
\begin{equation}
    \vec{\theta}\rightarrow \vec{\theta} +  \vec{\delta\theta},
\end{equation} 
where $\vec{\delta\theta}$ is a vector of random values sampled from a Gaussian distribution with zero mean and standard deviation $\sigma=0.01$ or $0.001$.
It imitates the effect of the over/under-rotations occurring in the circuit preparation.
This random deviation is not added for each iteration in the classical optimization part of ADAPT-VQE to keep the simulation fast. This leads to an underestimate of the error impact on ADAPT-VQE.
From Fig.~\ref{fig:rot_error}, we see that the \NewName\ suffers moderately from the rotational error, while ADAPT-VQE shows a high sensitivity to the rotational error where the energy starts to increase after a certain number of iterations.
It is because the parameters in ADAPT-VQE are first tightly optimized to an optimal point, and the energy is sensitive to any perturbation from the optimal point. 
Since ADAPT-VQE goes through the VQE subroutine at every adaptive step, each from the optimized parameters from the last step, the rotational error leads to a worse initial point for the next round of classical optimization. 
In contrast, perturbations in the update parameters have less impact in the non-variational approach since a range of parameter values can lead to energy reduction. 
The number of function evaluations required by noisy simulations of ADAPT-VQE to reach the same energy accuracy is thus larger than the \NewName. 
Finally, the randomized adaptive method is almost unchanged under the same level of rotation error, because the energy error it can attain is already larger than the noise level in the simulation.

\begin{figure}
    \centering
    \begin{subfigure}
        \centering
        \includegraphics[width=0.45\textwidth]{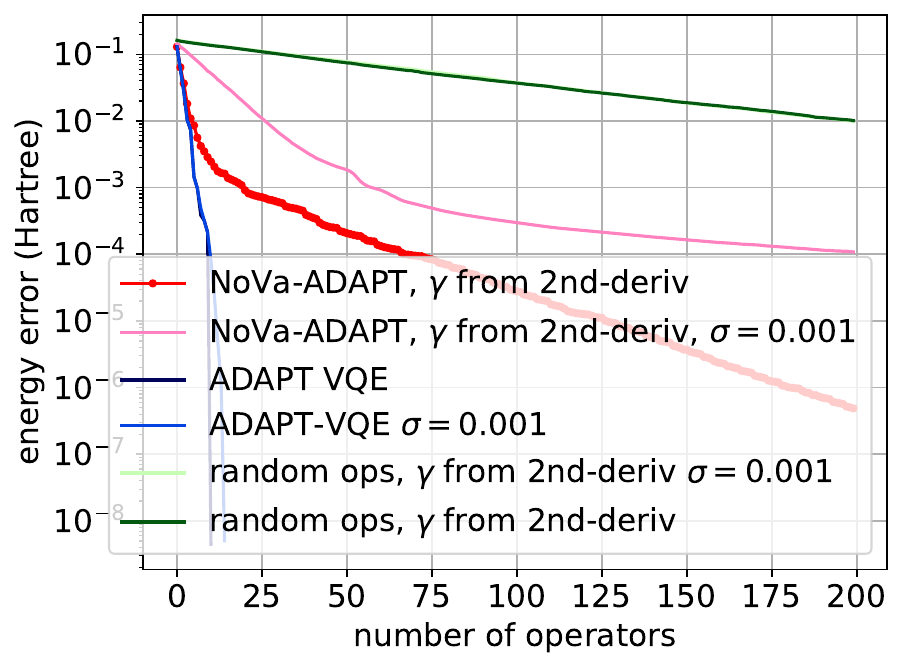}
    \end{subfigure}
    \begin{subfigure}
        \centering
        \includegraphics[width=0.45\textwidth]{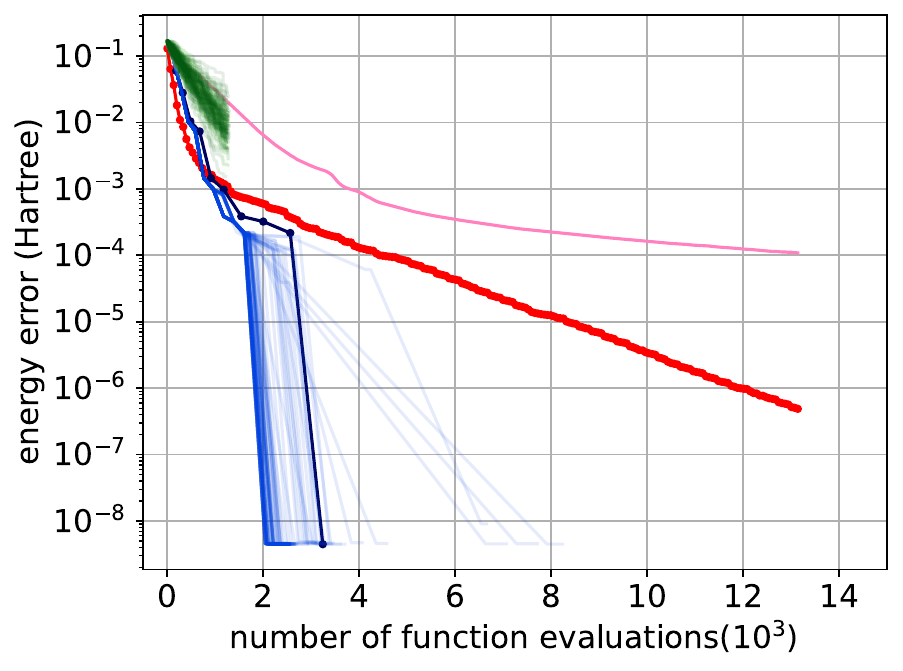}
    \end{subfigure}
    \caption{Results for $H_4$ molecule of bond length $1.5$\AA\ with gradient measurement errors. Energy error is plotted as a function of the number of operators added to the state (upper panel), and as a function of the number of function evaluations (lower panel). Blue curves show the results of ADAPT-VQE. Red curves show the results of the \NewName\ with $\gamma$ computed from the second derivative. Green curves show the results of randomly picked operators from the pool with $\gamma$ computed from the second derivative. All results shown here are based on 100 circuit realizations, either as individual realizations or as the average.}
    \label{fig:grad_error}
\end{figure}

The gradient measurement error is inserted to the simulation by adding a random value to the calculated first derivative of the pool operators during the operator selection procedure.
\begin{equation}
    \left.\frac{\partial E_{A_i}}{\partial \theta}\right|_{\theta=0}\rightarrow \left.\frac{\partial E_{A_i}}{\partial \theta}\right|_{\theta=0} +  \delta_i.
\end{equation}
where $\delta_i$ is a random value sampled from Gaussian distribution with zero mean and standard deviation $\sigma$.
This form of error captures the effect of statistical uncertainty associated with repeated measurements.
It will affect the choice of operators for ADAPT-VQE and \NewName.
Since the \NewName\ and the randomized algorithm also use the measured gradient to estimate the update parameter, they are affected by the gradient error in two ways. 
From Fig.~\ref{fig:grad_error}, we see that the \NewName\ is affected moderately by the gradient error, while ADAPT-VQE is robust against the gradient error, since a less-than-optimal choice of operator can still produce a decent variational ansatz. 
We remark that gradient errors are not introduced in the gradient-based classical optimization part of ADAPT-VQE, so the energy error in the noisy ADAPT-VQE simulation is underestimated.
Meanwhile, the randomized adaptive method is also unchanged under the same level of gradient error.

\section{ADAPT-VQE---NoVa-ADAPT hybrid approach} \label{sec:hybrid}

\begin{figure}
    \centering
    \begin{subfigure}
        \centering
        \includegraphics[width=0.45\textwidth]{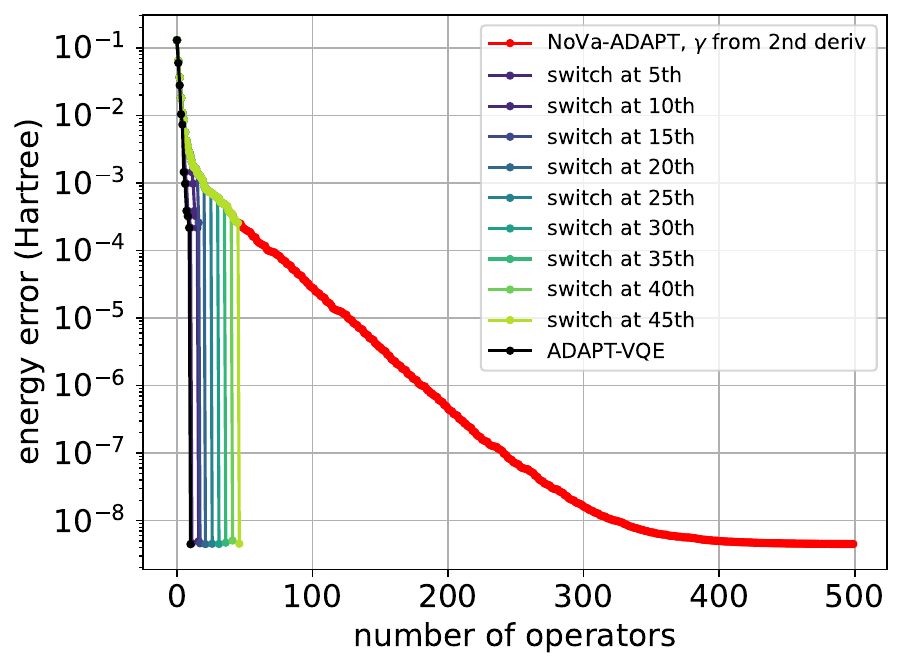}
    \end{subfigure}
    \hfill
    \begin{subfigure}
        \centering
        \includegraphics[width=0.45\textwidth]{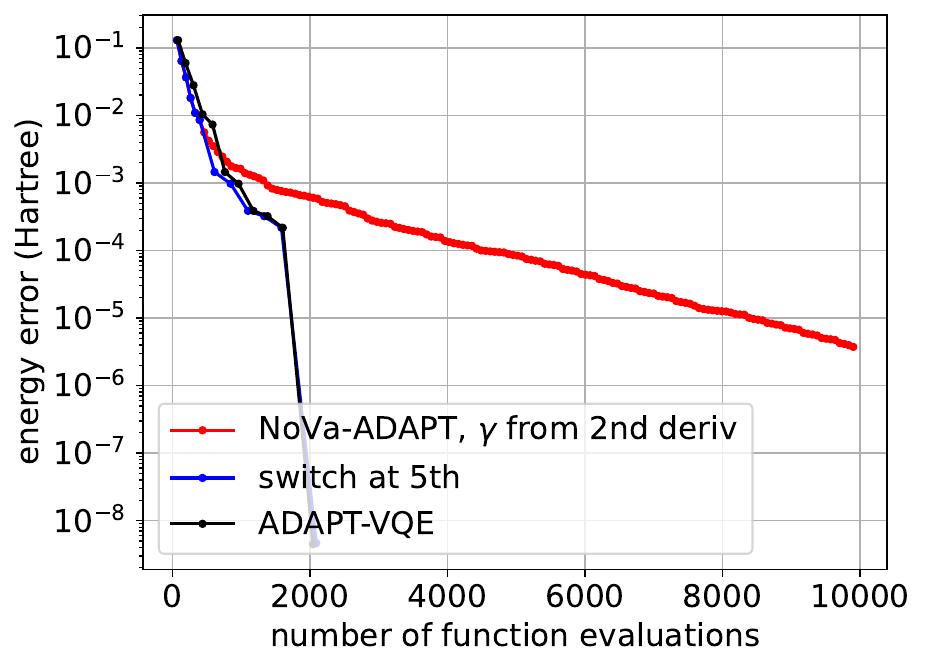}
    \end{subfigure}
    \caption{Results for $H_4$ molecule of bond length $1.5$\AA\ using the \NewName\ (red), ADAPT-VQE (black), and procedures starting with the non-variational update and switching to ADAPT-VQE at the 5, 10, 15, 20, 25, 30, 35, 40, 45-th iteration (see legend). Energy error is plotted as a function of the number of operators in the state (upper panel), and as a function of the number of function evaluations (lower panel). }
    \label{fig:switch}
\end{figure}

While the \NewName\ can potentially lower the measurement cost, it adds more operators to the state and leads to a deeper quantum circuit. 
To compensate for this drawback on the circuit depth and leverage the desired features of each algorithm, we explore the strategy of combining ADAPT-VQE and the \NewName. 
In Fig.~\ref{fig:hf}(b), we observe that the \NewName\ performs better at the early stage of the algorithm, offering an energy reduction comparable to that of ADAPT-VQE.
Based on this observation, we start with the \NewName\ and then switch to ADAPT-VQE after a certain number of iterations.
In Fig.~\ref{fig:switch}(b), we see that for the $\text{H}_4$ Hamiltonian with a bond distance of $1.5$ \AA, switching at the 5th iteration (blue) can slightly reduce the number of function evaluations required to reach chemical accuracy.

\section{Conclusions} \label{sec:conclusion}
We introduce the \NewName\ for preparing ground states, which both selects the operators and estimates the coefficients for updating the state based on the energy gradient information, without performing optimization.
Through numerical simulation, we show that the \NewName\ performs comparably to ADAPT-VQE for a small molecular Hamiltonian in terms of measurement cost at the early stage of the algorithm, but has more operators in the circuit.
By getting rid of the classical optimization, the \NewName\ may provide more robustness against errors in circuit parameters due to imperfect control or gate synthesis. It appears more prone to gradient measurement errors than ADAPT-VQE, although the impact of gradient measurement errors in the latter is underestimated in our simulation. Further work is required to determine whether noisy measurement favors the variational approach, and whether the level of noise plays a role.
By choosing the most energy-relevant operators, the \NewName\ saves quantum and classical resources compared to previous feedback-based quantum algorithms that either include all operators in the Hamiltonian or use random operators.
In the case of a sub-optimal initial state, we find that both the non-variational ADAPT approach and the approach of adding randomized operators perform better than ADAPT-VQE in terms of measurement cost.
We also explore the strategy of combining ADAPT-VQE and the non-variational approach, which may provide a realistic way to construct the ground state using limited resources.

\begin{acknowledgements}
    This work was supported by NSF grant No. 2231328.  A.B.M. acknowledges support from Sandia National Laboratories’ Laboratory Directed Research and Development Program under the Truman Fellowship. Sandia National Laboratories is a multimission laboratory managed and operated by National Technology \& Engineering Solutions of Sandia, LLC, a wholly owned subsidiary of Honeywell International Inc., for the U.S. Department of Energy’s National Nuclear Security Administration under contract DE-NA0003525. This paper describes objective technical results and analysis. Any subjective views or opinions that might be expressed in the paper do not necessarily represent the views of the U.S. Department of Energy or the United States Government. SAND2024-15486O.

\end{acknowledgements}

\appendix

\section{Simulated Hamiltonian in qubit operations} \label{app:H}

We use Jordan-Wigner transformation to map the Hamiltonian for the $\text{H}_4$ molecule in the STO-3G basis with a bond distance of $1.5$ \AA\ to:
\begin{widetext}
\begin{equation}
\begin{split}
H=&- 0.92094 \\
&- 0.03974  X_0 X_1 Y_2 Y_3 
  -0.00906  X_0 X_1 Y_2 Z_3 Z_4 Z_5 Z_6 Y_7 
- 0.00906  X_0 X_1 X_3 Z_4 Z_5 X_6 
  -0.02877  X_0 X_1 Y_4 Y_5  \\
&- 0.02749  X_0 X_1 Y_6 Y_7 
  +0.03974  X_0 Y_1 Y_2 X_3  
+ 0.00906  X_0 Y_1 Y_2 Z_3 Z_4 Z_5 Z_6 X_7 
  -0.00906  X_0 Y_1 Y_3 Z_4 Z_5 X_6  \\
&+ 0.02877  X_0 Y_1 Y_4 X_5 
  +0.02749  X_0 Y_1 Y_6 X_7  
+ 0.02080  X_0 Z_1 X_2 X_3 Z_4 X_5 
  +0.02080  X_0 Z_1 X_2 Y_3 Z_4 Y_5  \\
&- 0.01998  X_0 Z_1 X_2 X_4 Z_5 X_6 
  -0.01161  X_0 Z_1 X_2 Y_4 Z_5 Y_6  
- 0.04000  X_0 Z_1 X_2 X_5 Z_6 X_7 
  -0.04000  X_0 Z_1 X_2 Y_5 Z_6 Y_7  \\
&- 0.00837  X_0 Z_1 Y_2 Y_4 Z_5 X_6 
  +0.02839  X_0 Z_1 Z_2 X_3 Y_4 Z_5 Z_6 Y_7  
+ 0.02001  X_0 Z_1 Z_2 X_3 X_5 X_6 
  -0.02839  X_0 Z_1 Z_2 Y_3 Y_4 Z_5 Z_6 X_7  \\
&+ 0.02001  X_0 Z_1 Z_2 Y_3 Y_5 X_6 
  +0.00650  X_0 Z_1 Z_2 Z_3 X_4  
- 0.00297  X_0 Z_1 Z_2 Z_3 X_4 Z_5 
  +0.00860  X_0 Z_1 Z_2 Z_3 X_4 Z_6  \\
&+ 0.01748  X_0 Z_1 Z_2 Z_3 X_4 Z_7 
  +0.00888  X_0 Z_1 Z_2 Z_3 Z_4 X_5 Y_6 Y_7  
- 0.00888  X_0 Z_1 Z_2 Z_3 Z_4 Y_5 Y_6 X_7 
  -0.00402  X_0 Z_1 Z_2 X_4  \\
&+ 0.01678  X_0 Z_1 Z_3 X_4 
  +0.01684  X_0 Z_2 Z_3 X_4  
+ 0.03974  Y_0 X_1 X_2 Y_3 
  +0.00906  Y_0 X_1 X_2 Z_3 Z_4 Z_5 Z_6 Y_7  \\
&- 0.00906  Y_0 X_1 X_3 Z_4 Z_5 Y_6 
  +0.02877  Y_0 X_1 X_4 Y_5  
+ 0.02749  Y_0 X_1 X_6 Y_7 
  -0.03974  Y_0 Y_1 X_2 X_3  \\
&- 0.00906  Y_0 Y_1 X_2 Z_3 Z_4 Z_5 Z_6 X_7 
  -0.00906  Y_0 Y_1 Y_3 Z_4 Z_5 Y_6  
- 0.02877  Y_0 Y_1 X_4 X_5 
  -0.02749  Y_0 Y_1 X_6 X_7  \\
&- 0.00837  Y_0 Z_1 X_2 X_4 Z_5 Y_6 
  +0.02080  Y_0 Z_1 Y_2 X_3 Z_4 X_5  
+ 0.02080  Y_0 Z_1 Y_2 Y_3 Z_4 Y_5 
  -0.01161  Y_0 Z_1 Y_2 X_4 Z_5 X_6  \\
&- 0.01998  Y_0 Z_1 Y_2 Y_4 Z_5 Y_6 
  -0.04000  Y_0 Z_1 Y_2 X_5 Z_6 X_7  
- 0.04000  Y_0 Z_1 Y_2 Y_5 Z_6 Y_7 
  -0.02839  Y_0 Z_1 Z_2 X_3 X_4 Z_5 Z_6 Y_7  \\
&+ 0.02001  Y_0 Z_1 Z_2 X_3 X_5 Y_6 
  +0.02839  Y_0 Z_1 Z_2 Y_3 X_4 Z_5 Z_6 X_7  
+ 0.02001  Y_0 Z_1 Z_2 Y_3 Y_5 Y_6 
  +0.00650  Y_0 Z_1 Z_2 Z_3 Y_4  \\
&- 0.00297  Y_0 Z_1 Z_2 Z_3 Y_4 Z_5 
  +0.00860  Y_0 Z_1 Z_2 Z_3 Y_4 Z_6  
+ 0.01748  Y_0 Z_1 Z_2 Z_3 Y_4 Z_7 
  -0.00888  Y_0 Z_1 Z_2 Z_3 Z_4 X_5 X_6 Y_7  \\
&+ 0.00888  Y_0 Z_1 Z_2 Z_3 Z_4 Y_5 X_6 X_7 
  -0.00402  Y_0 Z_1 Z_2 Y_4  
+ 0.01678  Y_0 Z_1 Z_3 Y_4 
  +0.01684  Y_0 Z_2 Z_3 Y_4  \\
&+ 0.11933  Z_0 
  +0.01684  Z_0 X_1 Z_2 Z_3 Z_4 X_5  
+ 0.01684  Z_0 Y_1 Z_2 Z_3 Z_4 Y_5 
  +0.10125  Z_0 Z_1  \\
&- 0.00839  Z_0 X_2 Z_3 Z_4 Z_5 X_6 
  -0.00839  Z_0 Y_2 Z_3 Z_4 Z_5 Y_6  
+ 0.05022  Z_0 Z_2 
  -0.01746  Z_0 X_3 Z_4 Z_5 Z_6 X_7  \\
&- 0.01746  Z_0 Y_3 Z_4 Z_5 Z_6 Y_7 
  +0.08996  Z_0 Z_3  
+ 0.06236  Z_0 Z_4 
  +0.09114  Z_0 Z_5  \\
&+ 0.07784  Z_0 Z_6 
  +0.10533  Z_0 Z_7  
- 0.02080  X_1 X_2 Y_3 Y_4 
  +0.02001  X_1 X_2 X_4 Z_5 Z_6 X_7  \\
&+ 0.02839  X_1 X_2 Y_5 Y_6 
  +0.02080  X_1 Y_2 Y_3 X_4  
+ 0.02001  X_1 Y_2 Y_4 Z_5 Z_6 X_7 
  -0.02839  X_1 Y_2 Y_5 X_6  \\
&- 0.04000  X_1 Z_2 X_3 X_4 Z_5 X_6 
  -0.04000  X_1 Z_2 X_3 Y_4 Z_5 Y_6  
- 0.01998  X_1 Z_2 X_3 X_5 Z_6 X_7 
  -0.01161  X_1 Z_2 X_3 Y_5 Z_6 Y_7  \\
&- 0.00837  X_1 Z_2 Y_3 Y_5 Z_6 X_7 
  +0.00888  X_1 Z_2 Z_3 X_4 X_6 X_7  
+ 0.00888  X_1 Z_2 Z_3 Y_4 Y_6 X_7 
  +0.00650  X_1 Z_2 Z_3 Z_4 X_5  \\
&+ 0.01748  X_1 Z_2 Z_3 Z_4 X_5 Z_6 
  +0.00860  X_1 Z_2 Z_3 Z_4 X_5 Z_7  
- 0.00297  X_1 Z_2 Z_3 X_5 
  +0.01678  X_1 Z_2 Z_4 X_5  \\
&- 0.00402  X_1 Z_3 Z_4 X_5 
  +0.02080  Y_1 X_2 X_3 Y_4  
+ 0.02001  Y_1 X_2 X_4 Z_5 Z_6 Y_7 
  -0.02839  Y_1 X_2 X_5 Y_6  \\
&- 0.02080  Y_1 Y_2 X_3 X_4 
  +0.02001  Y_1 Y_2 Y_4 Z_5 Z_6 Y_7  
+ 0.02839  Y_1 Y_2 X_5 X_6 
  -0.00837  Y_1 Z_2 X_3 X_5 Z_6 Y_7  \\
&- 0.04000  Y_1 Z_2 Y_3 X_4 Z_5 X_6 
  -0.04000  Y_1 Z_2 Y_3 Y_4 Z_5 Y_6  
- 0.01161  Y_1 Z_2 Y_3 X_5 Z_6 X_7 
  -0.01998  Y_1 Z_2 Y_3 Y_5 Z_6 Y_7  \\
&+ 0.00888  Y_1 Z_2 Z_3 X_4 X_6 Y_7 
  +0.00888  Y_1 Z_2 Z_3 Y_4 Y_6 Y_7  
+ 0.00650  Y_1 Z_2 Z_3 Z_4 Y_5 
  +0.01748  Y_1 Z_2 Z_3 Z_4 Y_5 Z_6  \\
&+ 0.00860  Y_1 Z_2 Z_3 Z_4 Y_5 Z_7 
  -0.00297  Y_1 Z_2 Z_3 Y_5  
+ 0.01678  Y_1 Z_2 Z_4 Y_5 
  -0.00402  Y_1 Z_3 Z_4 Y_5  \\
&+ 0.11933  Z_1 
  -0.01746  Z_1 X_2 Z_3 Z_4 Z_5 X_6  
- 0.01746  Z_1 Y_2 Z_3 Z_4 Z_5 Y_6 
  +0.08996  Z_1 Z_2  \\
&- 0.00839  Z_1 X_3 Z_4 Z_5 Z_6 X_7 
  -0.00839  Z_1 Y_3 Z_4 Z_5 Z_6 Y_7  
+ 0.05022  Z_1 Z_3 
  +0.09114  Z_1 Z_4  \\
&+ 0.06236  Z_1 Z_5 
  +0.10533  Z_1 Z_6  
+ 0.07784  Z_1 Z_7 
  -0.03517  X_2 X_3 Y_4 Y_5  \\
&- 0.02944  X_2 X_3 Y_6 Y_7 
  +0.03517  X_2 Y_3 Y_4 X_5  
+ 0.02944  X_2 Y_3 Y_6 X_7 
  -0.02174  X_2 Z_3 X_4 X_5 Z_6 X_7  \\
&- 0.02174  X_2 Z_3 X_4 Y_5 Z_6 Y_7 
  +0.01055  X_2 Z_3 Z_4 Z_5 X_6  
- 0.01865  X_2 Z_3 Z_4 Z_5 X_6 Z_7 
  +0.00330  X_2 Z_3 Z_4 X_6  \\
&- 0.01844  X_2 Z_3 Z_5 X_6 
  +0.00261  X_2 Z_4 Z_5 X_6  
+ 0.03517  Y_2 X_3 X_4 Y_5 
  +0.02944  Y_2 X_3 X_6 Y_7  \\
&- 0.03517  Y_2 Y_3 X_4 X_5 
  -0.02944  Y_2 Y_3 X_6 X_7  
- 0.02174  Y_2 Z_3 Y_4 X_5 Z_6 X_7 
  -0.02174  Y_2 Z_3 Y_4 Y_5 Z_6 Y_7  \\
&+ 0.01055  Y_2 Z_3 Z_4 Z_5 Y_6 
  -0.01865  Y_2 Z_3 Z_4 Z_5 Y_6 Z_7  
+ 0.00330  Y_2 Z_3 Z_4 Y_6 
  -0.01844  Y_2 Z_3 Z_5 Y_6  \\
&+ 0.00261  Y_2 Z_4 Z_5 Y_6 
  +0.07128  Z_2  
+ 0.00261  Z_2 X_3 Z_4 Z_5 Z_6 X_7 
  +0.00261  Z_2 Y_3 Z_4 Z_5 Z_6 Y_7  \\
&+ 0.09406  Z_2 Z_3 
  +0.05893  Z_2 Z_4  
+ 0.09410  Z_2 Z_5 
  +0.06483  Z_2 Z_6  \\
&+ 0.09428  Z_2 Z_7 
  +0.02174  X_3 X_4 Y_5 Y_6  
- 0.02174  X_3 Y_4 Y_5 X_6 
  +0.01055  X_3 Z_4 Z_5 Z_6 X_7  \\
&- 0.01865  X_3 Z_4 Z_5 X_7 
  -0.01844  X_3 Z_4 Z_6 X_7  
+ 0.00330  X_3 Z_5 Z_6 X_7 
  -0.02174  Y_3 X_4 X_5 Y_6  \\
&+ 0.02174  Y_3 Y_4 X_5 X_6 
  +0.01055  Y_3 Z_4 Z_5 Z_6 Y_7  
- 0.01865  Y_3 Z_4 Z_5 Y_7 
  -0.01844  Y_3 Z_4 Z_6 Y_7  \\
&+ 0.00330  Y_3 Z_5 Z_6 Y_7 
  +0.07128  Z_3  
+ 0.09410  Z_3 Z_4 
  +0.05893  Z_3 Z_5  \\
&+ 0.09428  Z_3 Z_6 
  +0.06483  Z_3 Z_7  
- 0.04234  X_4 X_5 Y_6 Y_7 
  +0.04234  X_4 Y_5 Y_6 X_7  \\
&+ 0.04234  Y_4 X_5 X_6 Y_7 
  -0.04234  Y_4 Y_5 X_6 X_7  
- 0.00689  Z_4 
  +0.09690  Z_4 Z_5  \\
&+ 0.05391  Z_4 Z_6 
  +0.09626  Z_4 Z_7  
- 0.00689  Z_5 
  +0.09626  Z_5 Z_6  \\
&+ 0.05391  Z_5 Z_7 
  -0.10062  Z_6  
+ 0.11281  Z_6 Z_7 
  -0.10062  Z_7
\end{split}
\end{equation}
\end{widetext}

\bibliography{database.bib}

\end{document}